\documentclass[reprint,aps,pre,showpacs,amsmath,superscriptaddress]{revtex4-1}

\usepackage{graphicx}
\usepackage{dcolumn}
\usepackage{bm}
\usepackage{color}
\usepackage{amssymb}

\newcommand{\kbt}[0]{ k_\mathrm{B}T}

\newcommand{\figref}[1]{Fig. #1}
\newcommand{\eqaref}[1]{Eq. (#1)}

\begin{document}

\title{Comparing kinetic Monte Carlo and thin-film modeling of transversal
instabilities of ridges on patterned substrates
}

\author{Walter Tewes}
\email{walter.tewes@uni-muenster.de}

\affiliation{%
 Institute for Theoretical Physics, University of M\"unster, Wilhelm-Klemm-Str.~9, 48149 M\"unster, Germany
}%

\author{Oleg Buller}%

\affiliation{%
 Institute for Physical Chemistry, University of M\"unster, Correnstr.~28/30, 48149 M\"unster, Germany
}%


\author{Andreas Heuer}
\affiliation{%
 Institute for Physical Chemistry, University of M\"unster, Correnstr.~28/30, 48149 M\"unster, Germany
}%
\affiliation{
  Center of Nonlinear Science (CeNoS), University of M\"unster, Corrensstr.~2, 48149 M\"unster, Germany
}%
\affiliation{
  Center for Multiscale Theory and Computation (CMTC), University of M\"unster, Corrensstr.~40, 48149 M\"unster, Germany
}
\author{Uwe Thiele}
\email{u.thiele@uni-muenster.de}
\affiliation{%
  Institute for Theoretical Physics, University of M\"unster, Wilhelm-Klemm-Str.~9, 48149 M\"unster, Germany
}%
\affiliation{
  Center of Nonlinear Science (CeNoS), University of M\"unster, Corrensstr.~2, 48149 M\"unster, Germany
}%
\affiliation{
  Center for Multiscale Theory and Computation (CMTC), University of M\"unster, Corrensstr.~40, 48149 M\"unster, Germany
}%
\author{Svetlana V. Gurevich}
\affiliation{%
  Institute for Theoretical Physics, University of M\"unster, Wilhelm-Klemm-Str.~9, 48149 M\"unster, Germany
}%
\affiliation{
  Center of Nonlinear Science (CeNoS), University of M\"unster, Corrensstr.~2, 48149 M\"unster, Germany
}%
\affiliation{
  Center for Multiscale Theory and Computation (CMTC), University of M\"unster, Corrensstr.~40, 48149 M\"unster, Germany
}%


%
 \begin{abstract}
We employ kinetic Monte Carlo (KMC) simulations and a thin-film
continuum model to comparatively study the transversal (i.e.,
Plateau-Rayleigh) instability of ridges formed by molecules on
pre-patterned substrates. It is demonstrated that the evolution of
the occurring instability qualitatively agrees between the two
models for a single ridge as well as for two weakly interacting
ridges. In particular, it is shown for both models that the
instability occurs on well defined length and time scales which are, for the KMC model,
significantly larger than the intrinsic scales of thermodynamic
fluctuations. This is further evidenced by the similarity of
dispersion relations characterizing the linear instability modes.

\end{abstract}

\maketitle


\section{\label{Introduction}Introduction}
Over the past two decades, effects of pre-structured substrates
on the wetting behavior of liquids on solid substrates have been extensively experimentally
investigated to achieve determined liquid structures or to control
the dynamic self-assembly of organic molecules that show a
liquid-like behavior. Thereby, the pre-structuring can be of
topographical type \cite{yoon2008nanopatterning,seemann2005wetting},
purely chemical
\cite{PhysRevLett.86.4536,sehgal2002pattern,zhang2003patterning,mukherjee2008control,konnur2000instability,gau1999liquid},
i.e., affecting the local wetting properties, or a combination of both
\cite{LMWC2012jcp}.

Theoretically and numerically, the behavior of the molecules can be
modeled, e.g., by \textit{kinetic Monte Carlo} (KMC) simulations,
\textit{Molecular Dynamics} (MD) simulations or by various
\textit{continuum models}. Static liquid structures on a
  substrate with a chemical stripe-like pre-pattern and their
  morphological changes are investigated in
  \cite{BrLi2002jap,BaDi2000pre,lenz1998morphological} based on the
  minimization of effective interface energies.  In
\cite{BaDi2000pre}, a spatially varying effective interface potential
is employed to model a pre-structure patch in two dimensions, the
investigation in \cite{BrLi2002jap} is conducted for three-dimensional
liquid structures on a single pre-structure stripe and two adjacent
pre-structure stripes.  The dynamics of a liquid on a
chemically pre-patterned substrate is investigated in
\cite{KaKS2000l,PhysRevLett.86.4536} by direct numerical simulations
of thin-film equations with spatially varying Derjaguin (or disjoining) pressures.
Similar equations are considered in
\cite{thiele2003modelling,mechkov2008stability}, where bifurcation
diagrams for static ridge-like states are determined and
their transversal stability is analyzed.
A combination of topographical and chemical pre-patterns is accounted
for in \cite{LMWC2012jcp}, where atomistic KMC simulations are
conducted. MD simulations for liquids on
chemically structured substrates are performed in \cite{KLRD2006pf}.

Recently, the formation of bulges and droplets on substrates
  with stripe-like pre-patterns \cite{wang2011high} and the
nucleation and growth of structures on substrates with different types
of pre-patterns \cite{wang2012area} were investigated in detail in
vapor deposition experiments.  The nucleation and growth process in
these experiments was modeled in terms of KMC simulations
\cite{Wang2016press}, while the bulge and droplet formation on
pre-patterned stripes was recently addressed with a mesoscopic
thin-film model \cite{honisch2015instabilities}.

Whereas KMC and MD simulations can incorporate more details of
the specific interactions between the deposited molecules as well as
between molecules and substrate, continuum models are able to
address much larger length and time scales. Further, with continuum
models one may (semi-)analytically analyse instabilities of the
liquid structures \cite{thiele2003modelling,king2006linear,mechkov2008stability,BKHT2011pre} and
provide experimentalists with general results. Given the
advantages and disadvantages of the different theoretical and
numerical methods, it is evident that a mapping between the methods
is of great interest.

Here, we qualitatively compare results obtained with KMC simulations
and with a thin-film model for the dynamics of the Plateau-Rayleigh
instability of liquid ridges formed on a substrate with chemical
stripe-like pre-pattern. The classical Plateau-Rayleigh
instability \cite{Egge1997rmp} refers to the surface tension-driven instability of
axisymmetric liquid columns, bridges \cite{Plateau1873} or jets
\cite{Sava1833ac,rayleigh1878instability}. However, in our case of liquid
ridges on solid substrates, the base states are not axisymmetric but
have in the case of partially wetting liquids a roughly parabolic
cross section.  Therefore stability considerations have to
incorporate wettability in addition to surface tension. In the case
of patterned substrates, a sufficient contrast can eventually lead to
the stabilization of the Plateau-Rayleigh instability of liquid
ridges
\cite{thiele2003modelling,mechkov2008stability,honisch2015instabilities}. Note
that although the stability may be determined based on a purely
energy-based argument, the
fastest growing wavelength of the instability results from the interplay
of dynamics and energetics, i.e., for its determination dynamical models have to be studied.

We demonstrate that although the continuum thin-film model results from a long-wave approximation
of the Stokes equation, it qualitatively provides a good continuum limit
description of the KMC simulations. In particular, we show that
ridge structures formed in the KMC simulations are subject to a
transversal (i.e., Plateau-Rayleigh) instability with well defined
time and length scales, well separated from the intrinsic scales of
thermodynamic fluctuations. Note that for an axisymmetric elongated
soft matter system without substrate (e.g., a nanowire),
Ref.~\cite{muller2002template} employs a KMC model to describe a
similar Plateau-Rayleigh instability.

\section{Modeling Approaches}
\subsection{Continuum Model}
A classical modeling approach for thin layers of liquids is based on the
\textit{thin-film} or \textit{lubrication}
approximation of the Stokes equation \cite{oron1997long}. The
resulting \textit{thin-film equation} is an evolution equation
for the local height $h=h(\mathbf{x},t),\mathbf{x}\in \Omega \subset\mathbb{R}^2$ of the
liquid. It corresponds to a conservation law and can be written in gradient dynamics form \cite{Mitl1993jcis,Thie2010jpcm}
\begin{align}
 \partial_t h =-\boldsymbol{\nabla}\cdot\mathbf{j}
= \boldsymbol{\nabla}\cdot\left[M(h)\boldsymbol{\nabla}\frac{\delta \mathcal{F}}{\delta h}\right].
  \label{thinfilmequation}
\end{align}
Here, $\mathcal{F}$ is an energy functional and $M(h)$ is a mobility
function, which depends on the boundary conditions employed for the
Stokes flow at the solid-liquid interface. For \textit{no-slip} conditions,
the mobility reads $M(h)=h^3/3\eta$, where $\eta$ is the dynamic viscosity of
the liquid. In \cite{honisch2015instabilities}, different types of
mobilities are discussed in detail. Macroscopically, the
dominant term in the energy functional $\mathcal{F}$ is the energy of the
free surface of the film
\begin{align}
 \mathcal{F}_{\mathrm{surface}}=\int\limits_{\Omega}\gamma~\mathrm{d}s\approx\mathrm{const}+\underbrace{\int\limits_{\Omega}\frac{\gamma}{2}(\boldsymbol{\nabla}h)^2 \mathrm{d}\mathbf{x}}_{\mathcal{F}_{\mathrm{Laplace}}},
\end{align}
where $\gamma$ is the liquid-gas interfacial tension and
$\mathrm{d}s=\sqrt{1+ (\boldsymbol{\nabla}h)^2}$ denotes an area element
of the free surface that in the last step we have approximated by its
long-wave form.

 Considering mesoscopic scales, this term is typically supplemented by the \textit{wetting potential}
 \begin{align}
  \mathcal{F}=\mathcal{F}_{\mathrm{Laplace}}+\int\limits_{\Omega}\mathcal{U}(h) \mathrm{d}\mathbf{x},
 \end{align}
 which exhibits a minimum at a physically small film height, referred
 to as the \textit{adsorption layer} (or precursor film) height
 \cite{pismen2001nonlocal}. The derivative of the local energy
 $\mathcal{U}(h)$ w.r.t.\ film height corresponds to the negative of
 the Derjaguin (or disjoining) pressure.
 In a macroscopic or mesoscopic hydrodynamic context, the influence of
 the wetting potential is restricted to regions of small film heights
 near contact lines. It represents a convenient way to model
 wettability in the case of partially wetting liquids and does also
 relieve the moving contact line singularity
 \cite{pismen2001nonlocal,savva2011dynamics}. In a thermodynamic
context, the wetting potential is also often referred to as the
\textit{effective interface} or \textit{binding potential} (e.g.,
\cite{schick1990liquids,hughes2015liquid,Diet1988}), whereas the entire energy
functional is sometimes called \textit{interface hamiltonian}. In our
continuum approach, a spatially modulated model wetting potential is
employed that models the influence of the chemical pre-pattern on the
liquid by defining different mesoscopic contact angles on distinct
areas.

The film height-dependent part of the wetting potential employed in
the present work combines long-range attractive van der Waals and short-range
repulsive interactions \cite{pismen2001nonlocal}. Both
contributions are equally modulated by a function $g(x)$, so that the
non-dimensional wetting potential reads:
\begin{align}
\mathcal{U}(h)=\left(-\frac{1}{2h^2}+\frac{1}{5h^5}\right)(1+\rho g(\mathbf{x})).\label{disjoining_potential}
\end{align}
Here, $\rho$ is the contrast between more and less wettable areas,
where smooth transition regions are modeled by a piecewise sigmoidal
function $g(\mathbf{x})$. Similar wetting potentials were used
in \cite{pismen2006asymptotic,mechkov2008stability} and most recently
in \cite{honisch2015instabilities}. Also other forms were applied to
model stripe geometries, see e.g.~\cite{thiele2003modelling,konnur2000instability}.
Similar to \cite{honisch2015instabilities}, we consider stripe-like
pre-patterns, thus, the modulation function is chosen as:
\begin{align}
 g(\mathbf{x})=g(x)=1.0+\sum\limits_{j=1}^{2n}(-1)^j\mathrm{tanh}((x-x_j)/l_s),
\end{align}
where $n$ is the number of stripes, $x_j$ are the positions of the
transition regions between domains of different wettability and $l_s$
is the steepness of the transition. For all $(1+\rho g(x)) >0$,
the wetting potential corresponds to the partially wetting
case. Static one-dimensional solutions at a fixed overall mass are
given by roughly parabolic droplets with a contact angle given by
$\theta_e\approx\sqrt{2|U(h_{p})|}$. The drops are smoothly
connected via a contact line region to the equilibrium adsorption
layer of height $h_{p}$ \cite{glasner2003coarsening}
(cf.~\figref{\ref{fig:Steady_Growth}} (a)).  Thus, in our case the modulation
factor $g(x)$ changes the local equilibrium contact angle, but
always within the partially wetting regime. In the
case of \eqaref{\ref{disjoining_potential}}, more and less wettable
regions correspond to $g\approx-1$ and $g\approx 1$,
respectively. In the two-dimensional case of extended ridges,
pinning of the contact lines at the transition regions between more and
less wettable areas may result in a stabilization of the ridge with
respect to transversal instabilities, this stabilization is encoded
in the energetics of the corresponding one-dimensional solution and
is, e.g., studied in Refs.~\cite{thiele2003modelling,mechkov2008stability}.  In the following, we investigate
systems containing either a single or two pre-pattern stripes, that
are symmetrically arranged with respect to the axis $x=L_x/2$. Due to
periodic boundary conditions, the specific geometry is specified by
the width $w$ of a single pre-pattern stripe and the distance $d$
between two stripes. All direct numerical simulations for the
thin-film model are initialized by appropriate parabolas on top of a
adsorption layer height film, positioned on the pre-pattern
stripes. The time simulations are conducted using a generic finite
element framework (DUNE PDELab,
\cite{bastian2008genericI,bastian2008genericII,bastian2010generic})
employing bilinear ansatz functions on a rectangular grid. We use
periodic boundary conditions in both spatial directions.  For the
time-stepping, an implicit second order Runge-Kutta algorithm
\cite{alexander1977diagonally} is employed. For the case of a single
ridge (\figref{\ref{fig:ONESTRIPE_INSTAB}}), the simulations are
performed on a domain of $(L_x,L_y)=(160,400)$ with
$(N_x,N_y)=(160,400)$ elements, for the case of two ridges
(\figref{\ref{fig:TwostripesDNS}}), we consider $(L_x,L_y)=(200,600)$
with $(N_x,N_y)=(200,600)$.

\subsection{KMC Model}\label{secKMCModel}
\label{sec:kmc_model}
The KMC system is modeled by a lattice gas on a three
dimensional cubic lattice with the lattice constant $a$. Every
lattice site is either filled or empty as indicated by three
occupation numbers: two for the substrate with chemical pre-pattern
and one for the fluid. In particular, a lattice site
${\bf i}=\{i_x, i_y, i_z\}$ can be occupied by fluid particles $p_{\bf i}$ or
substrate sites $s_{\bf i}$ and $g_{\bf i}$, representing the
less and more wettable regions, respectively. Since we are only
interested in the dynamics of the film, the occupation numbers for
the substrate and more wettable stripe sites stay fixed. The Hamiltonian is written as
\begin{equation}
\label{hamiltonian}
\begin{split}
  H = -\epsilon_{pp} \; \frac 1 2 \sum_{{\bf i,j}} f(r_{\bf {ij}}) p_{\bf i} p_{\bf j} \qquad \\
  - \epsilon_{pg} \; \frac 1 2 \sum_{{\bf i,j}} f(r_{\bf ij}) p_{\bf i} g_{\bf j} - \epsilon_{ps} \; \frac 1 2 \sum_{{\bf i,j}} f(r_{\bf ij}) p_{\bf i} s_{\bf j},
\end{split}
\end{equation}
where the $\epsilon_{xy} \, (x,y \in \{ p,g,s \})$ are the interaction
parameters and $f(r_{\bf ij})$ is a scaling function that depends on
the distance $r_{\bf ij}$ between the particles on positions ${ \bf i}$
and ${\bf j}$. It is defined as follows:
  \begin{center}
    \begin{tabular}{l|c|c|c|c|c}
      \toprule
      $r_{\bf ij}/a$& 0.0 & 1.0 & $\sqrt 2$ & $\sqrt 3$ & $ > \sqrt 3$ \\ \hline
      $f(r_{\bf ij})$       & 0.0 & 1.0 & 1.0       & 0.5 & 0.0  \\
    \end{tabular}
  \end{center}
Interactions up to the third nearest neighbors (common corner) are
taken into account. As interaction parameters, we choose two parameter
sets corresponding to two different temperatures:
\begin{center}
    \begin{tabular}{l|c|c|c}
      \toprule
       set      &  $ \epsilon_{pp}/ \kbt$  &  $\epsilon_{ps}/\kbt $    &   $\epsilon_{pg}/\kbt$  \\  \hline
       P1       & 1.0000 & 0.5000 & 0.7000    \\
       P2       & 0.6250 & 0.3125 & 0.4375   \\
    \end{tabular} 
\end{center}
where $k_{\mathrm{B}}$ is the Boltzmann constant and $T$ the
temperature. The movement of the fluid particles is realized by Kawasaki
dynamics \cite{Kawasaki1972}. The binding energy of a particle is
compared to the energy it would have on a randomly chosen site from
the six nearest neighbor positions. The move to this position is
accepted according to the standard Metropolis criterion
\cite{Metropolis53}. The overall MC time step $\Delta t$ is
represented by all fluid particles performing a MC move.

\begin{figure}[t]
  \centering
  \includegraphics[width=0.45\textwidth]{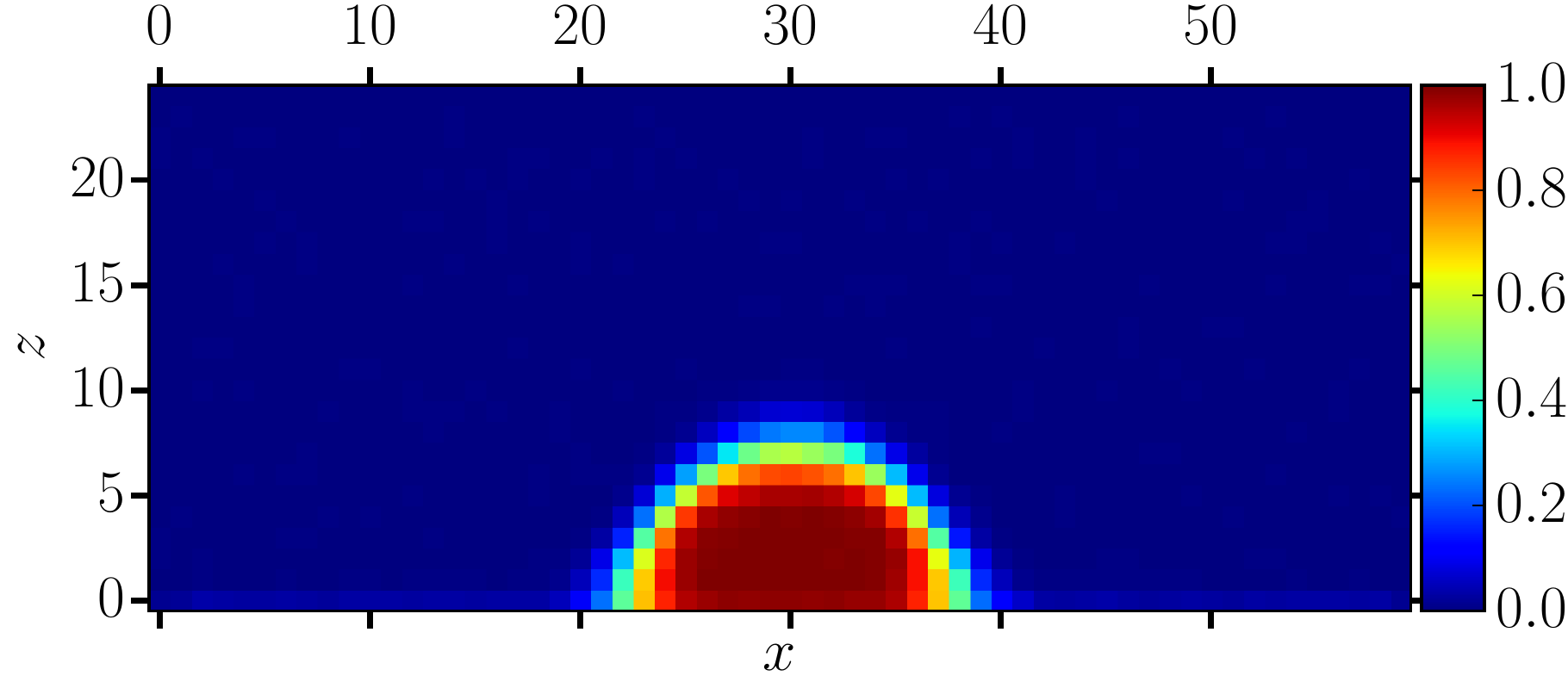}
  \caption{\label{fig:dprofMC} Density profile of the pseudo 2D droplet used to create the ridge
    in the MC system for the parameter set P2.}
\end{figure}

To set up the initial ridge geometry for a single stripe, we create a
two dimensional droplet profile in {$x$-} and {$z$-}direction which
is then extended in the $y$-direction. Thereby, we proceed as follows:
A simulation box of the size of $(60\times 25) a^2$ with 100 fluid
particles is set up. In the downwards {$z$-direction}, the
fluid is confined by
a fixed monolayer of substrate sites $s_{\bf i}$ at $i_z = 0$,
in the upwards direction by a hard wall, such that all attempts to move to sites with $i_z \ge 25$
are rejected.
In {$x$-direction}, periodic boundary
conditions are implemented.  In order to extend the system in
{$y$-direction}, likewise periodic boundary conditions are used. As
only one grid point plane in {$y$-direction} is considered,
this effectively results in self-interaction of the particles. After equilibration, a droplet
is formed. By shifting the center of mass of the droplet to the center
of the simulation box and averaging over 2000 independent
realizations, a well defined density profile (see
Fig. \ref{fig:dprofMC}) is created that is later used as a probability
map.
This particle density profile is used to generate the initial
ridge of length $L_y$ oriented in the $y$-direction. For every value
of the coordinate $y=0,.., L_y-1$, an individual configuration is
generated by randomly assigning a particle number on $x$ and $z$
positions following the probabilities as given by the density
profile. By this procedure, a well defined realization of a ridge on
the lattice is formed. The pre-structure is represented by more
attractive sites $g_{\bf i}$, which are incorporated into the substrate
plane where they replace the substrate sites $s_{\bf i}$ in a region of
width $w$ at the center of the domain. This pattern is then
continued in the $y$-direction thereby forming a more wettable
stripe of width $w$. The initial ridge is created centered on this
stripe in the same way as described above. Note that for
both parameter sets, the fluid is partially wetting, in particular,
the pre-patterned stripe has a smaller contact angle than the bare substrate.
All simulations are then
repeated multiple times to obtain a statistical description.

\section{Transversal Instabilities\label{Transversal Instabilities}}

\subsection{Single Ridge on Pre-patterned Substrate}
As a model system which is of great relevance for experiments, we
study the instability of a single ridge formed on top of a pre-pattern
stripe.  On homogeneous substrates, such ridge solutions are always
unstable with respect to transversal modulations (see
\cite{mechkov2007contact,thiele2003modelling}). As shown, e.g., in
\cite{honisch2015instabilities}, the pre-pattern can stabilize the
structure for certain regions of parameter space spanned by the
pre-pattern strength, pre-pattern size and the liquid volume in the
ridge. Here, we focus on the regime where the ridge is transversally
unstable. However, the broken translational invariance of the
substrate in $x$-direction simplifies the investigations of the large
scale behavior through the KMC simulations, since translational
fluctuations of the ridge are reduced.

\begin{figure*}[t]
\centering
\includegraphics[width=.50\textwidth]{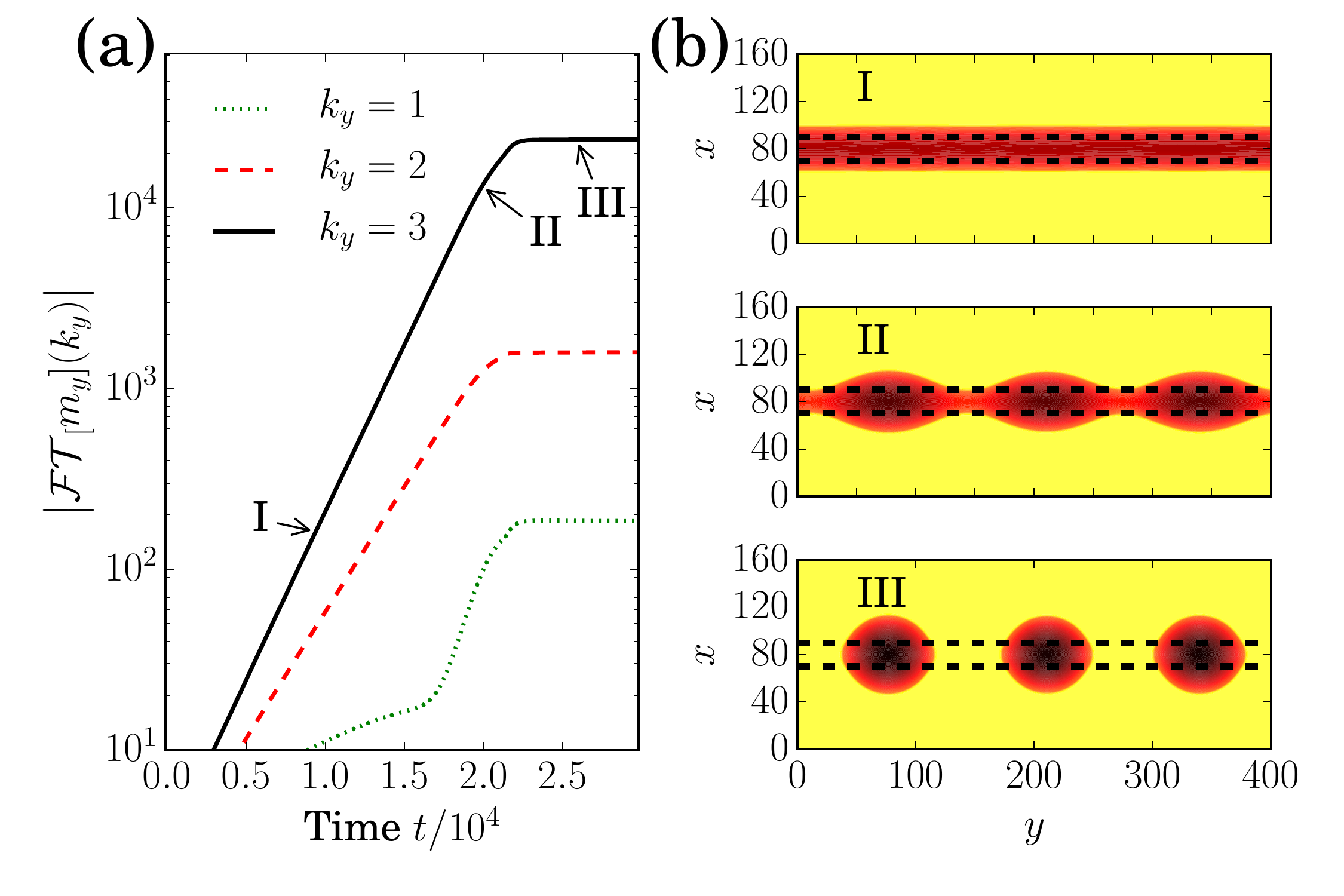}~\includegraphics[width=.50\textwidth]{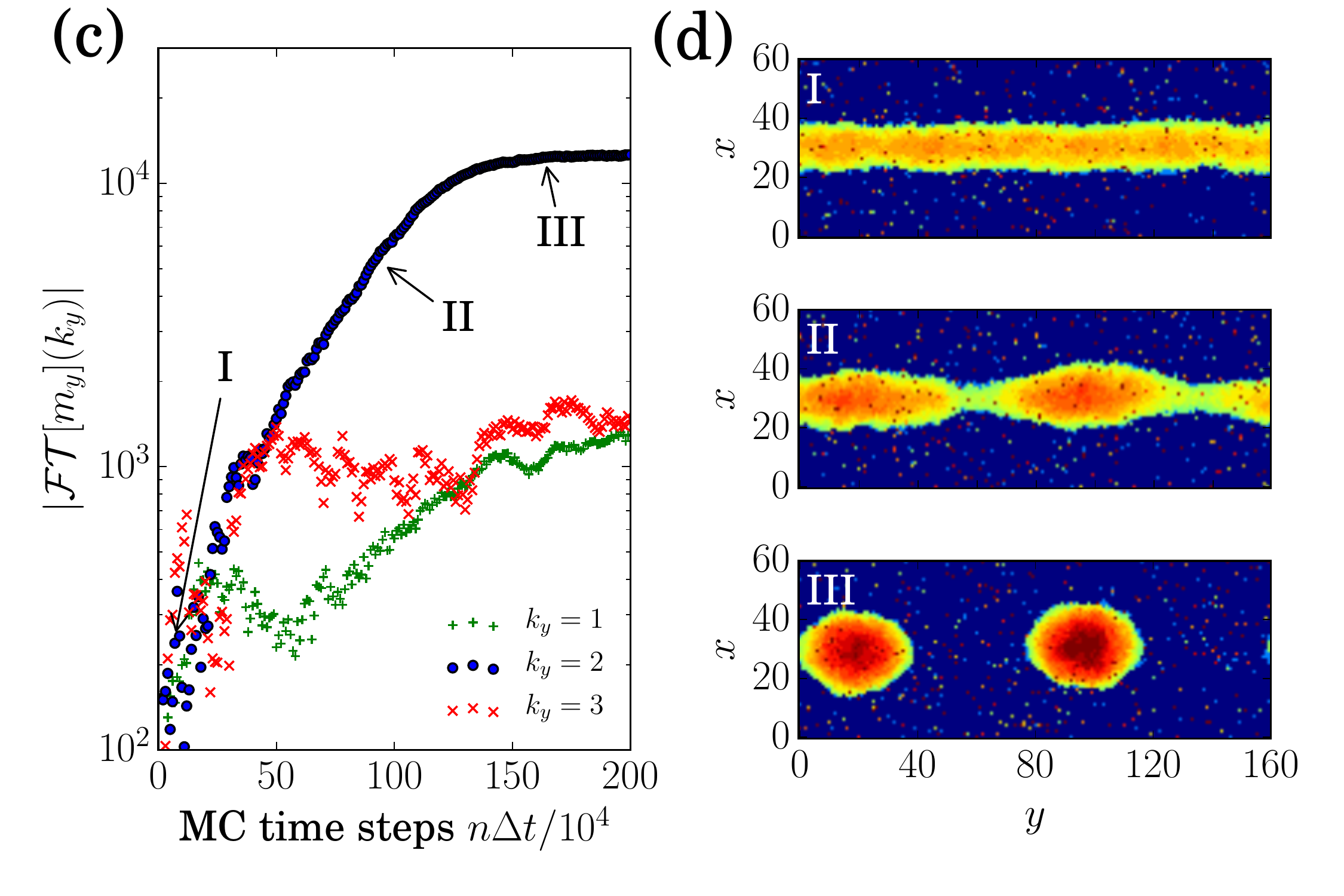}
\caption{\label{fig:ONESTRIPE_INSTAB}  (a) Plot of the absolute value of the Fourier transform $\left|\mathcal{FT}[m_y]\right|$ (modes $k_y=1,2,3$) over time in logarithmic presentation for the
    evolution of the transversal instability of a single ridge in a
    DNS of the thin-film model
    \eqaref{\ref{thinfilmequation}}. Corresponding snapshots are given
    in (b). The mean film height is $h_0=2.25$, further system
    parameters are $\rho=0.2$, $l_s=3.0$ and the stripe width
    $w=20$. The dashed lines in (b) visualize the borders of the more
    wettable region.  (c) Analogue plot of the absolute value of the Fourier transform (modes $k_y=1,2,3$) over time $n\Delta t$ for the evolution
    of the transversal instability of a single ridge obtained in a KMC
    simulation for parameter set P2, a system size of $L_y = 160a$ and
    $w = 10a$.  (d) Corresponding snapshots.}
\end{figure*}

In \figref{\ref{fig:ONESTRIPE_INSTAB} (b),(d)}, the transversal instabilities
and the resulting bulges on top of the pre-pattern are shown
for the two different models. In addition, as a measure for the volume
of molecules that are redistributed along the ridge, in
\figref{\ref{fig:ONESTRIPE_INSTAB} (a),(c)}, we show for each model the
time-evolution of the amplitudes of the first harmonic modes of the
field integrated in $x$- and $z$-direction. For the discrete KMC model, these quantities
are determined as the absolute values of the discrete Fourier transform of the
occupation field summed over $x$ and $z$ at a given transversal wavenumber $k_y$. It reads:
\begin{align}
\left|\mathcal{FT}[m_y](k_y,t)\right| = & \left | \sum_{n_y=0}^{L_y-1} m_y(n_y,t) e^{-2\pi i \frac{n_y k_y}{L_y}}  \right |, \\
 m_y(n_y,t) = & \sum_{n_x, n_z} P(n_x, n_y, n_z),
\end{align}
where $P(n_x, n_y, n_z)$ is the occupation field of the ridge.  For
the continuum model, the analogue of these quantities
is defined through the continuous Fourier transform of the height profile
integrated over $x$:
\begin{align}
  \left|\mathcal{FT}[m_y](k_y,t)\right|&=\left|\int m_y(y,t)e^{-2\pi i\frac{k_yy}{L_y}}\mathrm{d}y\right|,\label{fourierdef}\\m_y(y,t)&=\int h(x,y,t)\mathrm{d}x. 
\end{align}
The absolute values of the Fourier transform are shown individually for the first harmonic modes of
re-distribution, in particular, for wavenumbers $k_y=1,2,3$.

For the continuum model, the instability develops through a
rather extended phase of exponential growth that corresponds to the
linear instability regime. As expected due to the non-deterministic
nature of the model, in the KMC simulations this regime is influenced
by noise. However, an exponential growth regime of the $k_y=2$ mode can
be clearly identified. After the initial regime of linear instability,
both models show a nonlinear phase in which the ridge decays into
separated droplets via pinch-off events. Although the system is
clearly in the nonlinear regime in the later stages of the pinch-off process,
the growth of $\left|\mathcal{FT}[m_y]\right|$ remains surprisingly exponential in the early stage of this process,
still approximately with the same growth rate as in the initial regime. A
surprisingly long-lasting linear behavior has been reported before
for the Cahn-Hilliard equation \cite{sander2000unexpectedly}.  The
transversal instability and its dependency on system parameters will
now be analyzed in more detail for both models.
\subsubsection{Analysis of the Instability for the Continuum Model}
In the case of the continuum model, the dispersion relation and the
shape of the unstable modes can be obtained in a standard linear
stability analysis that results in an eigenvalue problem which is
solved numerically.

Denoting a one-dimensional steady state (stable or unstable) by
$h_0(x)$, a linear stability analysis w.r.t.\ transversal instability
modes is performed employing the ansatz
\begin{align}
 h(x,y,t)=h_0(x)+h_1(x)e^{\beta t+2\pi i\frac{k_yy}{L_y}}.
\end{align}
Note that $k_y$ in the exponential ansatz is scaled by $2\pi/L_y$, so
that its definition coincides with the one used in \eqaref{\ref{fourierdef}}.
Then, $2\pi k_y/L_y$ corresponds to the typically employed system size-independent
wavenumber. As described in more detail in
\cite{honisch2015instabilities,thiele2003modelling}, this ansatz combined with a
linearization of \eqaref{\ref{thinfilmequation}} leads to the eigenvalue problem
\begin{align}
 \mathcal{L}[h_0(x),k_y]h_1(x)=\beta h_1(x),\label{eigenvalueproblem}
\end{align}
where the eigenvalue $\beta$ of the linear operator $\mathcal{L}[h_0(x),k_y]$ corresponds to the growth rate of
transversal modulations with the wavenumber $k_y$.
The linear operator depends on the mobility function ($M(h)$
in \eqaref{\ref{thinfilmequation}}). Although the stability
criterion can be obtained by purely energetic considerations based
on capilarity and wettability, the specific form of
the dynamical equation influences the dispersion relation.
In particular, to obtain the fastest growing unstable mode, the full
dynamical equation has to be considered. Droplets formed in accordance
with the resulting wavelength might exist for a long transient before
successive coarsening \cite{honisch2015instabilities}. In realistic experiments,
the final post-coarsening equilibrium state of a single drop  is often
not reached, i.e.\ the dynamical pathway is of interest.

\begin{figure}[h]
\centering
\includegraphics[width=.39\textwidth]{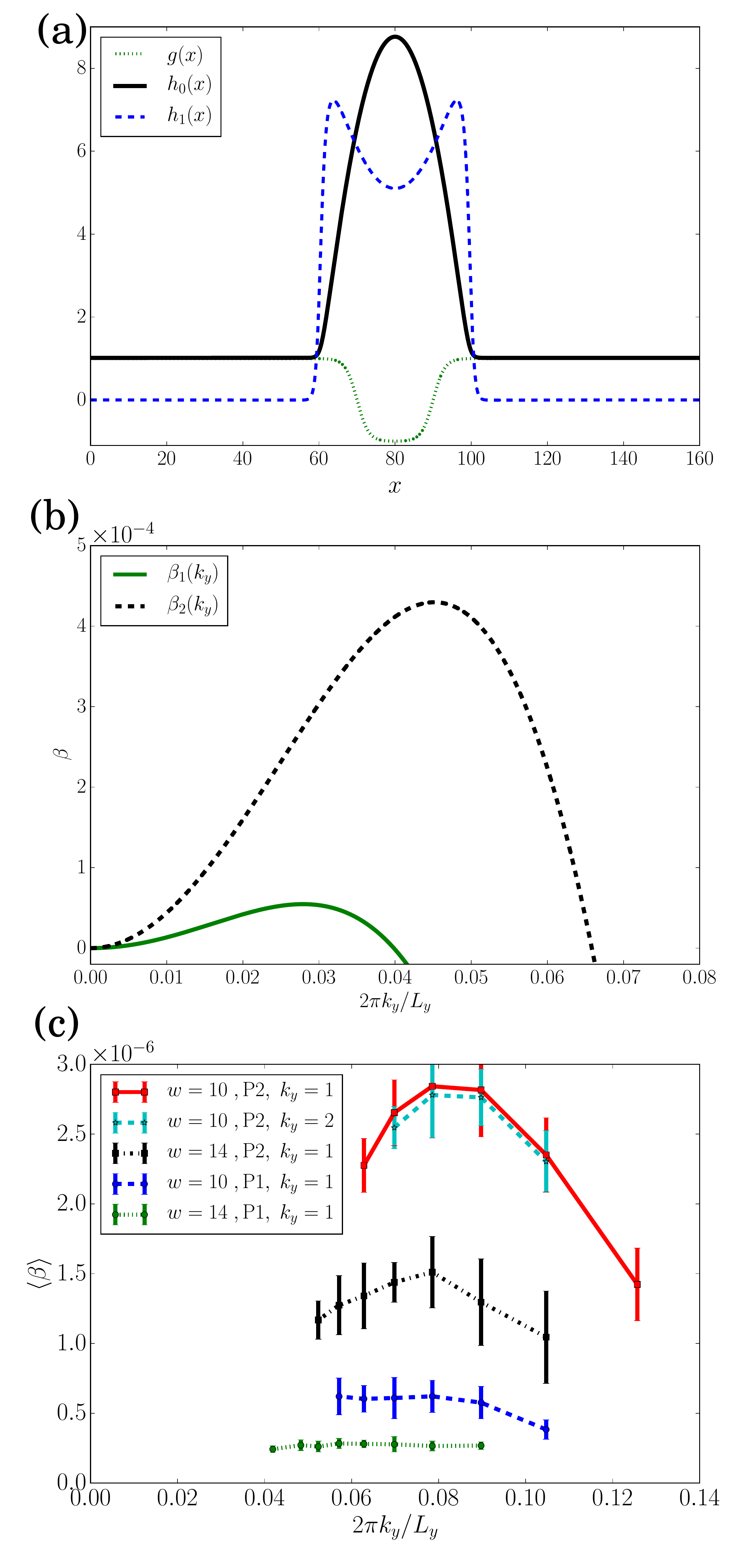}
\caption{\label{fig:Steady_Growth}
    (a) For the thin-film model, we show a one-dimensional steady
    state $h_0(x)$ (ridge cross-section), pre-pattern function $g(x)$
    and unstable transversal mode $h_1(x)$ at
    $k_{y,\mathrm{max}}$. The parameters are $h_0=2.25$, $\rho=0.2$,
    $w=20$.  (b) \label{fig:Dispersion} Dispersion relations of the
    transversal instability of a single ridge for two different
    pre-pattern stripe widths.  For $\beta_1$ and $\beta_2$,
    we chose $w=40$ and $w=20$, respectively. All other parameters are
    as above.  (c) Dispersion relations extracted from the MC
    simulations, P1 and P2 refer to the parameter sets given in section
    \ref{secKMCModel}.  The growth rate is determined by fitting the
    exponential regime of $\left|\mathcal{FT}[m_y]\right|$ for the selected
    mode. The errorbars represent the standard
    deviation of the averaging procedure.}
\end{figure}

In \figref{\ref{fig:Steady_Growth} (a)}, we show the one-dimensional
steady state solution $h_0$ as well as the eigenfunction $h_1$ for the
fastest growing wavenumber. The parameters correspond to the ones of
the direct numerical simulations (DNS) in \figref{\ref{fig:ONESTRIPE_INSTAB} (a),(b)}.
By numerical continuation techniques, the unstable eigenvalue and the
corresponding eigenfunction are followed in $k_y$ to obtain the
transversal dispersion relation (see
\figref{\ref{fig:Dispersion}}(b)). The dispersion relation is
reminiscent of a fourth order polynomial but can only be described
approximately by such a function due to the dependence of the
eigenfunction on $k_y$ that results in  contributions that are higher order in $k_y$.
At $k_y=0$, the eigenfunction $h_1$ has a bimodal form and corresponds
to the neutral \textit{growth mode}. The growth rate is zero for
$k_y=0$, as imposed by the mass conserving dynamics of the form
$\partial_t h=-\nabla \cdot \mathbf{J}$ of
\eqaref{\ref{thinfilmequation}} and the non-vanishing integral of the
eigenfunction $h_1$. We show the dispersion relation for the stripe
width employed in the DNS and for another, larger stripe width where
eventually, for larger values of $\rho$ than the one shown in
\figref{\ref{fig:Dispersion}} (b), pinning of the contact line
and stabilization of the ridge occurs.
\subsubsection{Analysis of the Instability for the KMC Model}
For the KMC model, the dispersion relations for the transversal
instability can only be determined by direct numerical
simulations. Furthermore, as already discussed at
\figref{\ref{fig:ONESTRIPE_INSTAB}}(c),(d), the initial, linear regime
of the instability in the KMC simulations is influenced by noise, such
that the linear growth rate is neither always present nor unique if
individual simulations are considered. Nonetheless, one may determine
the growth rate in a statistical sense. To do so, we consider systems
of different longitudinal size $L_y$ and determine
$\left|\mathcal{FT}[m_y]\right|$ for the dominant mode, compare
\figref{\ref{fig:ONESTRIPE_INSTAB} (c)}.  In analogy to the dispersion
relation obtained with the continuum model, for small values of $L_y$,
the modes with $k_y=1$ dominate. Increasing $L_y$, eventually higher
modes become dominant what implies that only a $k_y$-range around the
maximum of the dispersion relation can be obtained. The results are
shown in \figref{\ref{fig:Steady_Growth} (c)}.  For $w=10$ and P2, the
results for the growth rate of the $k_y=1$ modes are confirmed by a
corresponding analysis for the second mode ($k_y=2$) using a system
twice as large.  For each system size at least 10 appropriate
$\left|\mathcal{FT}[m_y]\right|$ curves are recorded and fitted by a
function of the form $f(t) = a e^{\beta t}$ for an appropriate time
range. Only fits with a maximum fit-error of 3\% are considered and
averaged in order to obtain the mean growth rate $\langle\beta\rangle$.
The results of this procedure are shown in
\figref{\ref{fig:Steady_Growth} (c)} for the two parameter sets and
pre-pattern stripe widths $w= 10$ and $w = 14$. As expected, only
growth rates around the maximum of the dispersion relation can be
extracted. For larger $L_y$, the instability is naturally dominated by
higher modes close to the maximum of the dispersion relation, i.e.,
small $k_y$ values can not be accessed by looking at the dominant
modes. For smaller $L_y$ (larger $k_y$) there is a sharp decrease
towards zero, which could not be better resolved with the present
statistics (compare to continuum model \figref{\ref{fig:Steady_Growth}
  (b)}).  Overall, we find that the dispersion relations extracted
from the KMC simulations qualitatively well agree with the ones
determined with the continuum model.  Note, however, that due to the
fluctuations in the system, lower growth rates are more difficult to
extract.

At higher temperatures (i.e., case P2), the maxima of the
dispersion relations are higher and can therefore be extracted more
precisely, even though the larger fluctuations imply larger variances.
In continuum models, an increased temperature can be reflected in
increased transport coefficients. In the particular thin-film model
 \eqaref{\ref{thinfilmequation}} used, this would correspond to a decreased
value of the viscosity $\eta$. Assuming that this effect dominates an
also possible effect on the wetting behavior (decreasing contact
angle with increasing temperature), one finds that the growth rate of
the instability which is inversely proportional to the viscosity,
increases with temperature. This is consistent with the corresponding
effect seen in the KMC simulations (\figref{\ref{fig:Steady_Growth}
  (c)}).
\subsection{Two Weakly Interacting Ridges on Pre-patterned Substrate}
\begin{figure*}[]
\centering
\includegraphics[width=.95\textwidth]{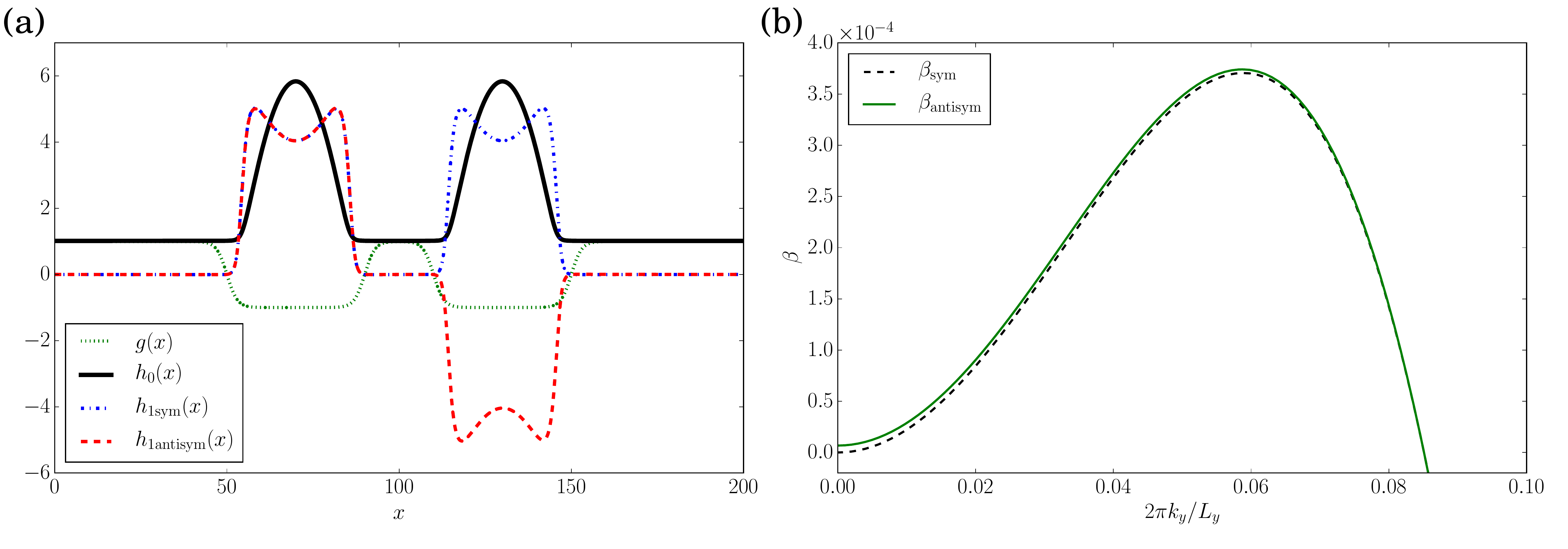}{}
\caption{\label{fig:Steady_Growth_Two}
 (a) For the thin-film model of the two-stripe system, we show a one-dimensional steady
  state $h_0(x)$ (cross-section of ridges), the pre-pattern function $g(x)$
  and the symmetrical and antisymmetrical unstable transversal modes $h_{1,\mathrm{sym}}(x)$ and
  $h_{1,\mathrm{antisym}}(x)$, respectively, both at
  $k_{y,\mathrm{max}}$. Parameters are $h_0=2.0$, $\rho=0.2$, $w=40$, $l_s=3$, $d=20$ (distance of the stripes).\\
  (b) \label{fig:DispersionTwo} Corresponding dispersion
  relations of the two unstable transversal instability modes. The dispersion
  relation of the symmetric mode approaches zero at $k_y=0$ due to
  mass conservation while the one of the antisymmetric mode
  approaches at $k_y=0$ a non-zero growth rate corresponding to the
  coarsening mode due to mass transfer of 1d drops, demonstrating the interaction of the ridges.}
\end{figure*}

The second pre-pattern geometry we consider is a system with two
neighboring more-wettable stripes. The distance $d$ between the stripes
does not correspond to the periodicity of the simulation domain
imposed by the boundary conditions. The question addressed by this
investigation is how the weak interaction of two liquid ridges on the
two stripes influences their instability. For the continuum model, we
perform a similar linear stability analysis as for the one-stripe
system above. For a symmetric two-ridge solution (one ridge on each
stripe, both identical), we
find two different transversal instability modes as shown
in \figref{\ref{fig:Steady_Growth_Two} (a)}.

The first mode, denoted by $h_{1,\mathrm{sym}}$, can be seen in analogy
to the instability mode for the one-stripe
system and proceeds through periodic transversal mass transfer (i.e.,
along each ridge) equally and synchronously in both ridges. The eigenvalue
of this mode goes to zero at $k_y=0$ as dictated by mass conservation,
compare \figref{\ref{fig:DispersionTwo}} (b).

The second, antisymmetric mode $h_{1,\mathrm{antisym}}$ shown in
\figref{\ref{fig:Steady_Growth_Two}} (a), corresponds to mass
transfer along each ridge \textit{and} between the two ridges in
such a way that the drop patterns developing on the two ridges are
shifted w.r.t.\ each other by half a period. This is a signature of
the weak interaction of the two ridges. Note that the growth rate of
the antisymmetric mode approaches a finite non-zero value at
$k_y=0$. In this case, the mode corresponds to the coarsening mode
due to mass transfer known for 1d drops
\cite{thiele2003modelling,Thie2007}.

\begin{figure*}
  \centering
  \includegraphics[width=0.4\textwidth]{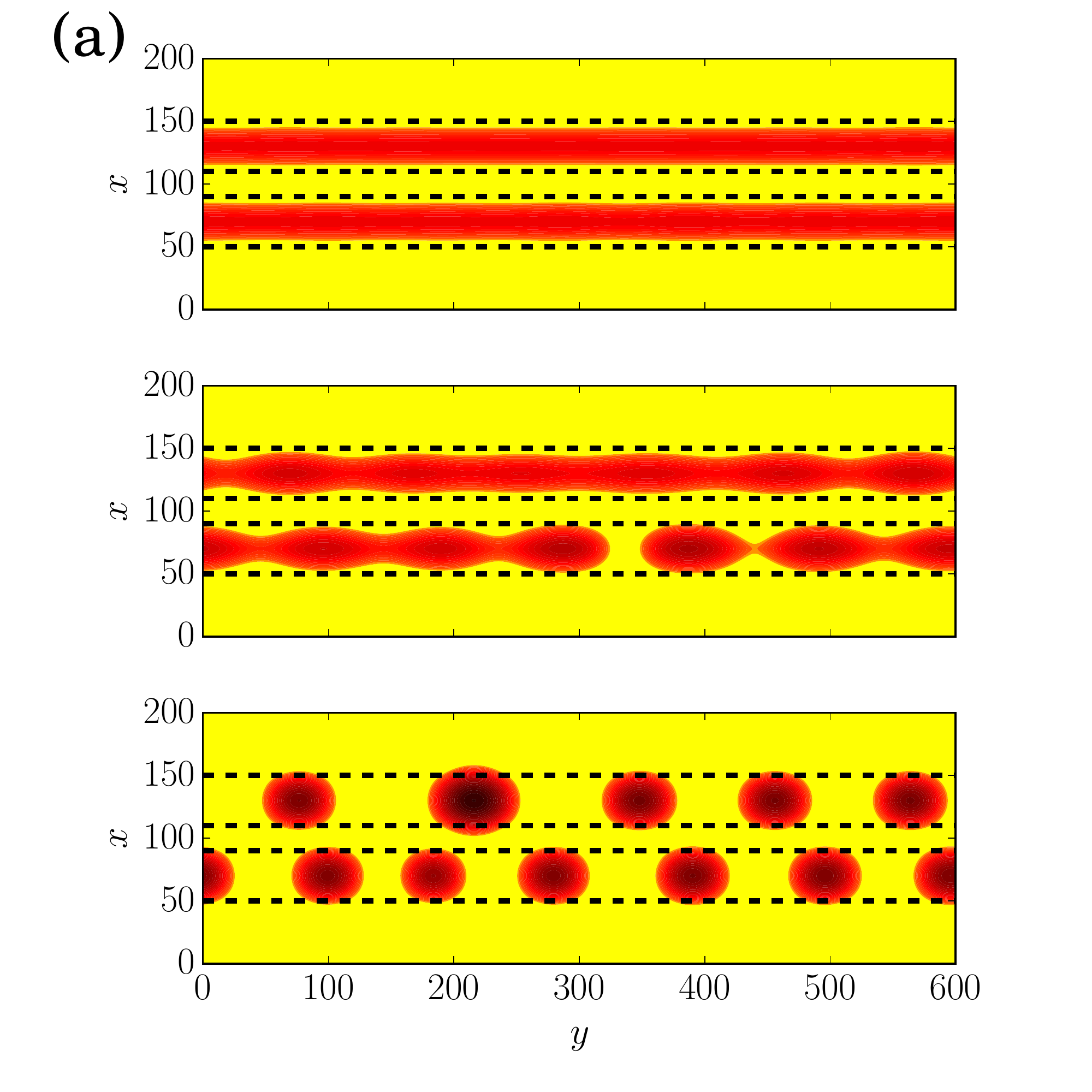}~\includegraphics[width=0.4\textwidth]{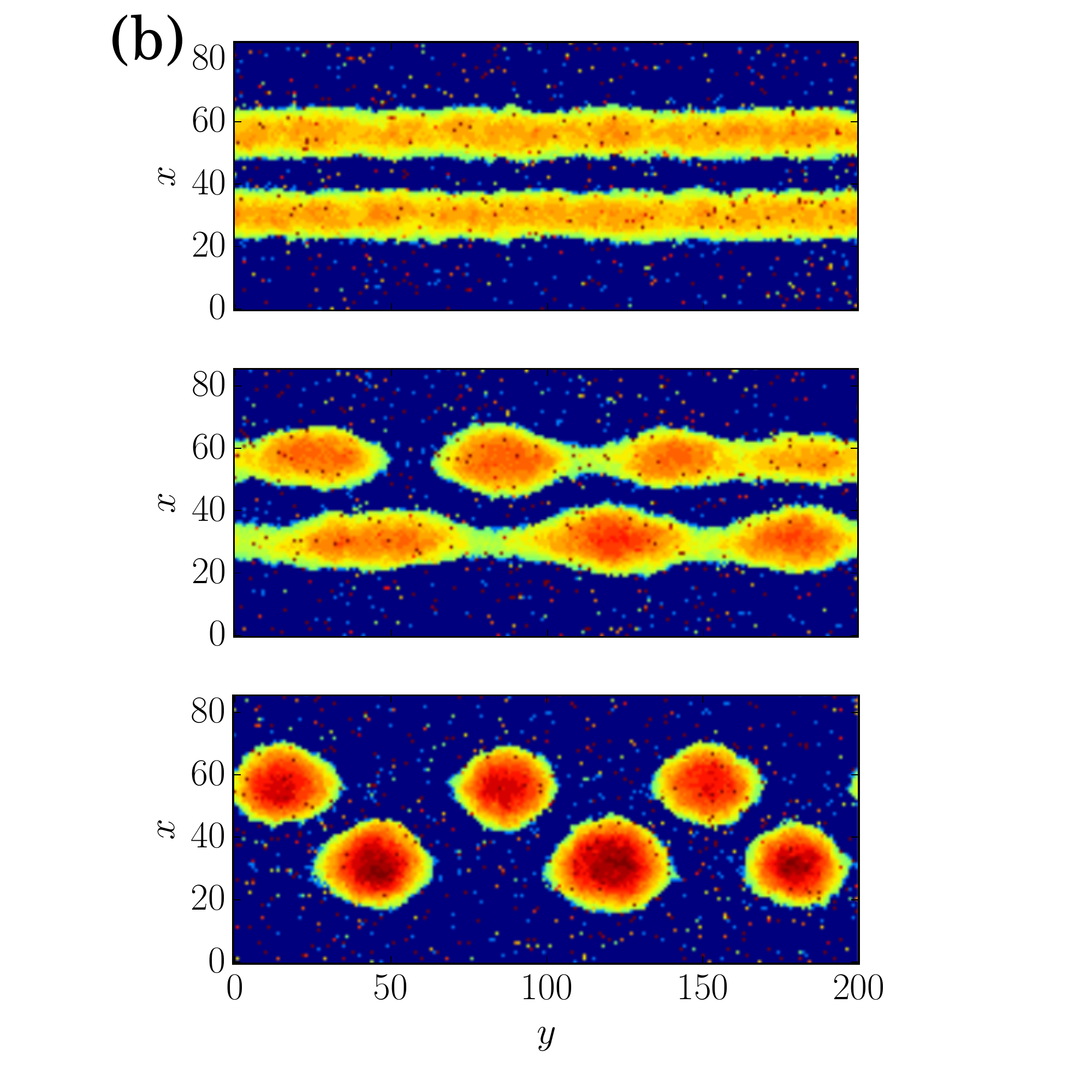}
  \caption{\label{fig:TwostripesDNS} {Snapshots from numerical
      simulations of the transversal instability of two weakly
      interacting ridges on pre-pattern stripes. (a) shows
      results for the thin-film model at parameters $h_0=2.0$,
      $\rho=0.2$, $w=40$, $d=20$. In (b), we give the
      corresponding results for the KMC simulations at parameters
      $w = 10a$, $d=16a$, $L_y=200a$ and set P2 of interaction constants.}}
\end{figure*}

For the dominant values of $k_y$, the growth rates of the two unstable
modes are almost identical, the one of the antisymmetric mode
  is only slightly larger. Therefore, the mode dynamically chosen by the
  system is arbitrary - it may be either of the two or some linear combination.
In the specific simulation of the continuum model, shown in \figref{\ref{fig:TwostripesDNS}} (a),
the system exhibits an instability dominated by the antisymmetric mode,
thus forming a configuration of alternating drops, which are subjected
to slow coarsening.

In the corresponding simulation with the KMC model shown in
\figref{\ref{fig:TwostripesDNS}} (b), one also observes a break up of the
two ridges in an alternating manner, thus also there the antisymmetric
instability mode dominates.

\section{Conclusion and outlook}
We have investigated the Plateau-Rayleigh instability of ridges of
molecules on pre-patterned substrates by means of two inherently
different modeling approaches, namely, a stochastic
discrete kinetic Monte Carlo (KMC) model and a
deterministic continuum hydrodynamic thin-film model.

We have systematically shown that despite the different nature of the
approaches, the results are in very good qualitative agreement. In
particular, the dynamics of the transversal ridge instabilities seen
with the KMC model on large time and length scales (compared to the
microscopic one of individual hopping events) is consistent with the
transversal instabilities as analyzed in the thin-film model: A
comparison of the Fourier analyses of typical time evolutions of
single ridges has shown that in both models one can identify
long regimes of exponential growth followed by droplet
pinch-off. The extended exponential regime has allowed us to extract
dispersion relations from the KMC model via a fitting and averaging
procedure. A comparison with dispersion relations obtained via
standard linear stability analysis within the thin-film model, i.e., by
the solution of the corresponding linear eigenvalue problem, has shown a
good qualitative agreement of the two approaches.  Likewise, we have
demonstrated a qualitative agreement of the evolution pathway in the
two-stripe system where in both models an antisymmetric droplet
pattern can evolve from the transversal instability. Therefore, also
there the linear stability result from the thin-film
model could be reproduced in the KMC simulations.

The focus of the present work has been on the qualitative
agreement of a stochastic discrete and a deterministic continuum
modeling approach in the context of a specific, experimentally
relevant system. However, we
would like to add a few remarks on the nature of the agreement
between the approaches and outline how a quantitative mapping may be
reached between averaged KMC simulations and a mesoscopic gradient
dynamics (or thin-film) model for the evolution of the film height profile.

As already emphasized in the introduction, to reach a
quantitative correspondence between the two models is a non-trivial
task, since the underlying physical assumptions about the system are
not identical. While the KMC model assumes a diffusive dynamics
with exclusively short-ranged particle interactions, the thin-film
model is derived from the standard equation for overdamped
hydrodynamics, namely the Stokes equation.

A first step in order to achieve a quantitative mapping between
the models would be to map the models in terms of their equilibrium
behavior, i.e., their statics/energetics. Here, one may consider a
mean field approximation of the KMC model in terms of a classical
lattice density functional theory (DFT) approximation. This could be
done for the case of different homogeneous substrates to then apply
the results to the heterogeneous case. In \cite{hughes2015liquid}, it
is shown how one is able to extract the interface tension and the
effective wetting potential from appropriate lattice DFT theories
with different interaction ranges (cf.~\cite{MaM2005jpm,TMTT2013jcp} for an alternative
approach via a Molecular Dynamics MD simulations). The resulting interface
Hamiltonian (or free energy) is then incorporated into a mesoscopic
continuum theory in gradient dynamics form and it is shown that
height profiles of mean field droplets obtained from the lattice DFT
are accurately reproduced. As explained above, the effective wetting
potential directly gives the Derjaguin (or disjoining) pressure that
appears in the thin-film equation.

Such a mapping on a purely energetic level does not only allow one
to investigate static ridge and droplet states but also to assess their
stability either via a second variation of the free energy or via a
dynamical approach. For the latter, one assumes that the dynamics
follows a gradient dynamics on the interface Hamiltonian. For
instance, for a thin-film equation like \eqaref{\ref{thinfilmequation}}, \cite{mechkov2007contact} shows
that the threshold of the transversal instability is entirely
encoded in the energy functional. To fully account for the dynamics
one needs to take a further step and extract from stochastic
discrete models as the KMC model the full mobility functions that
enter the thermodynamic fluxes in the gradient dynamics form of the
thin-film type model \eqaref{\ref{thinfilmequation}}.  This could be done
by a direct numerical fitting process or analytically, starting from
a Cahn-Hilliard type dynamical mean field model such as the one
discussed in \cite{monson2008mean}. However, the employed stochastic
model(s) should allow for the extraction of all effective transport
parameters, i.e., diffusion constant, slip length and viscosity to
fully account for transport by diffusion at very small
(sub-monolayer) film height and by advection at larger film
height. A discussion of these transport modes in the context of a
thin-film evolution equation with a general polynomial mobility
function may be found in \cite{YSTA2016arxiv}.\\

\begin{acknowledgments}
This work was supported by the Deutsche Forschungsgemeinschaft within the
Transregional Collaborative Research Center TRR 61.\\
W.T. and O.B. contributed equally to this work.
\end{acknowledgments}

\end{document}